\documentclass[conference,a4paper]{IEEEtran} 
\IEEEoverridecommandlockouts

\usepackage{cite}
\usepackage{amsmath,amssymb,amsfonts}
\usepackage{graphicx}
\usepackage{textcomp}
\usepackage{xcolor}
\usepackage{acro}
\usepackage{url}
\usepackage{siunitx}
\usepackage{booktabs}
\usepackage{tabularx}
\usepackage{multirow}
\usepackage{todonotes}
\usepackage{subfigure}
\usepackage{algorithmicx}
\usepackage{algorithm}
\usepackage[noend]{algpseudocode}
\usepackage{mathtools}

\usepackage{scalerel}
\usepackage{tikz}
\usetikzlibrary{svg.path}
\definecolor{orcidlogocol}{HTML}{A6CE39}
\tikzset{
  orcidlogo/.pic={
    \fill[orcidlogocol] svg{M256,128c0,70.7-57.3,128-128,128C57.3,256,0,198.7,0,128C0,57.3,57.3,0,128,0C198.7,0,256,57.3,256,128z};
    \fill[white] svg{M86.3,186.2H70.9V79.1h15.4v48.4V186.2z}
                 svg{M108.9,79.1h41.6c39.6,0,57,28.3,57,53.6c0,27.5-21.5,53.6-56.8,53.6h-41.8V79.1z M124.3,172.4h24.5c34.9,0,42.9-26.5,42.9-39.7c0-21.5-13.7-39.7-43.7-39.7h-23.7V172.4z}
                 svg{M88.7,56.8c0,5.5-4.5,10.1-10.1,10.1c-5.6,0-10.1-4.6-10.1-10.1c0-5.6,4.5-10.1,10.1-10.1C84.2,46.7,88.7,51.3,88.7,56.8z};
  }
}
\newcommand\orcidicon[1]{\href{https://orcid.org/#1}{%
\mbox{\begin{tikzpicture}[yscale=-1,transform shape, scale=0.03] 
\pic{orcidlogo};
\end{tikzpicture}}}}

\usepackage{hyperref}
\hypersetup{
    colorlinks = true,
    allcolors=blue,
    urlcolor=black,
    linkcolor=black
}

\usepackage{xspace}
\let\OldTexttrademark\texttrademark
\renewcommand{\texttrademark}{\OldTexttrademark\xspace}%

\usepackage{fancyhdr}
\fancypagestyle{IEEEtitlepagestyle}{
    \fancyhf{} 
     
    \fancyhead[C]{
        \small This work has been published at the \emph{1st Workshop on Resilient Networks and Systems (ReNeSys)}, Sept. 2025, Ilmenau/Germany under the Creative Commons Attribution 4.0 International License (CC BY 4.0).
        \textit{Digital Object Identifier \href{http://dx.doi.org/10.14279/depositonce-24399}{10.14279/depositonce-24399}}
        }
}

\DeclareSIUnit{\kbps}{k\acl{bit}/s}
\DeclareSIUnit{\mbps}{M\acl{bit}/s}
\DeclareSIUnit{\gbps}{G\acl{bit}/s}
\DeclareSIUnit{\tbps}{T\acl{bit}/s}
\DeclareSIUnit{\B}{\acl{byte}}
\DeclareSIUnit{\kB}{k\acl{byte}}
\DeclareSIUnit{\mB}{M\acl{byte}}
\DeclareSIUnit{\gB}{G\acl{byte}}
\DeclareSIUnit{\tB}{T\acl{byte}}
\DeclareSIUnit{\b}{\acl{bit}}
\DeclareSIUnit{\kb}{k\acl{bit}}
\DeclareSIUnit{\mb}{M\acl{bit}}
\DeclareSIUnit{\gb}{G\acl{bit}}
\DeclareSIUnit{\tb}{T\acl{bit}}
\DeclareSIUnit{\mpps}{Mpps}

\newcolumntype{Y}{>{\centering\arraybackslash}X}

\def\BibTeX{{\rm B\kern-.05em{\sc i\kern-.025em b}\kern-.08em
    T\kern-.1667em\lower.7ex\hbox{E}\kern-.125emX}}

\newcommand\citeN\cite

\newcommand\fig[1]{Figure~\ref{fig:#1}}

\newcommand{\figeps}[3][]{%
   \begin{figure}[tb!]
      \begin{center}
         \leavevmode
         \parbox[t]{#1}{%
            \resizebox{#1}{!}{\includegraphics{figures/#2}}
         }
         \caption{#3\vspace{-0.2cm}}
         \label{fig:#2}
      \end{center}
      \vspace{-0.45cm}
   \end{figure}
}

\newboolean{makevspace}
\newcommand{\cvspace}[1]{%
   \ifthenelse
   {\boolean{makevspace}}
   {\vspace{#1}}
   {}%
}
\acsetup{single}

\DeclareAcronym{TG}{
    short = TG,
    long = traffic generator,
}

\DeclareAcronym{RTT}{
    short = RTT,
    long = round-trip time,
}

\DeclareAcronym{IAT}{
    short = IAT,
    long = inter-arrival time,
}

\DeclareAcronym{MAT}{
    short = MAT,
    long = match-action table,
}

\DeclareAcronym{CBR}{
    short = CBR,
    long = constant bit-rate,
}

\DeclareAcronym{TCAM}{
    short = TCAM,
    long = ternary content addressable memory,
}

\DeclareAcronym{byte}{
    short = B,
    long = byte,
}

\DeclareAcronym{bit}{
    short = b,
    long = bit,
}

\begin{document}

\title{Enhancements to P4TG: Histogram-Based RTT Monitoring in the Data Plane}

\author{
  \IEEEauthorblockN{Fabian Ihle$^{\orcidicon{0009-0005-3917-2402}}$, Etienne Zink$^{\orcidicon{0009-0001-0879-535X
          }}$, and Michael Menth$^{\orcidicon{0000-0002-3216-1015}}$}
  \IEEEauthorblockA{University of T\"ubingen, Chair of Communication Networks
    \\Email: \{fabian.ihle, etienne.zink, menth\}@uni-tuebingen.de}
  \thanks{The authors acknowledge the funding by the Deutsche Forschungsgemeinschaft (DFG) under grant ME2727/3-1.}
}

\maketitle

\begin{abstract}
Modern traffic generators are essential tools for evaluating the performance of network environments.
P4TG is a P4-based traffic generator implemented for Intel Tofino switches that offers high-speed packet generation with fine-grained measurement capabilities.
However, P4TG samples time-based metrics such as the round-trip time (RTT) in the data plane and collects them at the controller.
This leads to a reduced accuracy.
In this paper, we introduce a histogram-based RTT measurement feature for P4TG.
It enables accurate analysis at line rate without sampling.
Generally, histogram bins are modeled as ranges, and values are matched to a bin.
Efficient packet matching in hardware is typically achieved using ternary content addressable memory (TCAM).
However, representing range matching rules in TCAM poses a challenge.
Therefore, we implemented a range-to-prefix conversion algorithm that models range matching with multiple ternary entries.
This paper describes the data plane implementation and runtime configuration of RTT histograms in P4TG.
Further, we discuss the efficiency of the ternary decomposition.
Our evaluation demonstrates the applicability of the histogram-based RTT analysis by comparing the measured values with a configured theoretical distribution of RTTs.
\end{abstract}


\vspace{-0.3cm}

\section{Introduction}
\label{sec:introduction}
With \acfp{TG}, the performance of a network environment can be evaluated by generating and measuring various traffic patterns.
Software-based \acp{TG} running on commodity hardware are more accessible, but offer lower performance compared to hardware-accelerated solutions~\cite{trex, EmGa17, CoVo24}.
In contrast, hardware-based \acp{TG} deliver higher performance but come at significantly higher cost.
The advent of the P4 language has enabled the development of affordable, programmable hardware solutions.
P4TG is a P4-based traffic generator implemented for the Intel Tofino\texttrademark\ switching ASIC~\cite{LiHae23, IhZi25, p4tg-git}.
It is more cost-efficient than commercial hardware \acp{TG} while supporting a broad range of protocols for traffic generation.
Further, P4TG offers extensive measurement capabilities.

The \ac{RTT} is a key metric for assessing network latency.
Significant variations in \ac{RTT} often signal network anomalies or instability in the network.
Timely detection of such deviations is essential for improving network resilience~\cite{HiCa16}.
In its initially published version, P4TG samples time-based statistics such as the \ac{RTT}.
This reduces the accuracy due to limited sampling.
Other P4-based \acp{TG}, like P4STA, sample the \ac{RTT} using an external monitoring host~\cite{KuSi20}.
This limits the accuracy to the sampling capabilities of the external host.
Sampling introduces bias in time-based statistics which can lead to inaccuracies in representing true RTT values.
Further, reliable anomaly detection is difficult with sampled values as outliers may be missed.

The contribution of this work is manifold.
We introduce a histogram-based \acs{RTT} measurement feature for P4TG that enables unsampled, line rate RTT measurement on Tofino\texttrademark 1 and 2.
Implementing histogram collection at line rate presents a challenge: networking hardware typically uses \ac{TCAM} to match packets against rules for classification and forwarding.
However, efficiently matching a value to a range in \acs{TCAM}, e.g., to assign it to a histogram bin, is a well-known and non-trivial problem~\cite{BrHa18, SuKi10, SuKi10_2, DoSu6}.
We address this challenge by applying a range-to-prefix conversion algorithm that encodes ranges using multiple ternary match entries.
The resulting histograms enable accurate and detailed \ac{RTT} distribution analysis, including precise calculations of the mean, standard deviation, and percentiles.
Because every \ac{RTT} is measured without sampling, the approach also enhances resilience testing by enabling more reliable detection of anomalies.
In this paper, we first describe the data plane implementation and runtime configuration of \ac{RTT} histograms in P4TG.
Next, we discuss the efficiency of the ternary decomposition, and finally, we demonstrate the applicability of the histogram-based \ac{RTT} measurement in an example network.
\section{Technical Background}
In this section, we give a brief introduction to the P4 programming language.
Then, we introduce the traffic generator P4TG.

\subsection{The P4 Programming Language}
P4 is a programming language used to implement custom data planes in network devices~\cite{BoDa14}.
For packet processing, so-called \acp{MAT} are applied.
A \ac{MAT} contains a composite key field of multiple header or metadata fields from the packet.
A packet is matched with its key field defined in the \acs{MAT}.
On matching an entry of the \acs{MAT}, an associated action is executed.
Those actions are also implemented in the data plane and define packet processing behavior, e.g., forwarding or header manipulation.
Additionally, a \acs{MAT} in P4 can be associated with a counter.
The counter counts the number of matched packets per entry.

Multiple matching types for keys in \acp{MAT} exist in P4, such as the exact, ternary, and range match.
With a ternary match, a value and a bitmask are configured in the \acs{MAT}.
The ternary entry matches if the bitwise AND operation of the packet's key field value and the configured mask is equal to the configured ternary value.
This enables wildcarding and aggregation, making ternary matches suitable for implementing prefix-based routing or class-based filtering.
Ternary entries are stored in \ac{TCAM} which is used for high-speed packet classification in switches~\cite{BrHa18, SuKi10}.
Finally, the range match type allows matching a packet to an interval.
Here, a lower and an upper bound of the range are configured.

\subsection{The Traffic Generator P4TG}
P4TG is a P4-based traffic generator implemented for the Intel Tofino\texttrademark switching ASIC~\cite{LiHae23, p4tg-git}.
P4TG leverages the internal traffic generation ports of the Intel Tofino\texttrademark to achieve a generation capacity of up to $10 \times \SI{400}{\gbps}$ with \ac{CBR} or Poisson traffic.
In a recent update, P4TG was extended with various protocols, test automation, and support for the Intel Tofino\texttrademark\ 2 switching ASIC~\cite{IhZi25}.

Generated traffic can be routed through a network and can be fed back to P4TG to measure various statistics.
For that purpose, generated traffic contains a UDP payload with sequence number and timestamping information.
Currently, P4TG measures \acp{IAT}, \acp{RTT}, L1 and L2 traffic rates, frame sizes, frame types, e.g., unicast or multicast, Ethernet types, e.g., VLAN or IPv4, packet loss, and out of order packets.
The RTT value (in ns) is calculated by subtracting the transmission and reception timestamps of P4TG's physical interfaces.

Statistics are gathered in different ways.
Time-based statistics, i.e., \ac{IAT} and \ac{RTT}, are stored on a per-port basis in a register of the data plane during packet processing.
Those are then exported to the control plane using small metadata messages known as digests.
The rate of those digest messages is metered to not overwhelm the control plane.
Therefore, time-based statistics are sampled and may lose accuracy.
The minimum, maximum, mean, and standard deviation of time-based statistics are calculated based on those sampled values.
Frame metrics, such as the size and type, are aggregated in the data plane and are periodically read by the control plane.
They are not sampled and correspond to the exact number of counted frames.
The statistics are collected in the control plane and are made available in a REST API endpoint for external use.
Further, the web-based frontend of P4TG leverages this endpoint to visualize the statistics in real-time.


\section{RTT Histogram Support in P4TG}
\label{sec:histogram}
In this section, we describe the implementation of the histogram feature for \ac{RTT} measurements in P4TG.
We begin with the general design in the data plane, followed by the modeling of histogram bins using range-to-prefix conversion.
Finally, we explain the runtime configuration of histograms and how they improve the accuracy of RTT statistics.

\subsection{Data Plane Histogram Design}
The histogram functionality of P4TG is implemented in the data plane which allows for line rate mapping of packet's \acp{RTT} to bins, i.e., no sampling is applied.
For this purpose, a \acs{MAT} which models the histogram is added to P4TG.
Entries in this \acs{MAT} correspond to bins of the histogram, i.e., to a specific time range.
All incoming packets are matched according to their \qty{32}{bit} wide \acs{RTT}.
The \acs{MAT} is associated with a counter that tracks the number of matched packets per entry.
The counter has a width of \qty{64}{bit} per entry and therefore can count trillions of packets without overflowing.
If a packet does not match to a bin, it is counted as an outlier.

\subsection{Bin Modeling with Range-to-Prefix Conversion}
For range matching, the P4 language provides the range matching type.
On the Intel Tofino\texttrademark, range matching is supported only for fields up to \qty{20}{\bit} wide due to \acs{TCAM} limitations.
Since the RTT field in P4TG is \qty{32}{bit}, we cannot apply native range matches directly.
While one possible workaround would be to extract a \qty{20}{\bit} subrange using bit shifting, this is not feasible in the P4TG pipeline due to internal hardware constraints.
Instead, we apply a range-to-prefix conversion algorithm~\cite{GuMc01} that allows efficient matching using multiple ternary entries.
The algorithm receives an integer range $[L, R]$ of a bin and decomposes it into a minimal set of prefixes (power-of-two aligned blocks) so that every integer in the range is covered.
It iteratively selects the largest prefix to the current lower bound that fits into the range.
The combination of those blocks results in full coverage for the range.
For example, the range $[4, 7]$ is covered by the prefix $\{01\!*\!*\}$ while the range $[3, 8]$ requires the prefixes $\{0011, 01\!*\!*, 1000\}$~\cite{GuMc01}.
Bins in a histogram are consecutive non-overlapping ranges.
The correctness and uniqueness of a solution under this condition has been proven by Sun~\cite{Su11}.



\subsection{Histogram Configuration}
The configuration of \acs{RTT} histograms is applied on a per-RX-port basis.
During runtime, histograms in P4TG are configurable with a minimum and maximum value, and the number of bins.
Bins can be defined with nanosecond precision.
Before traffic generation starts, those parameters can be configured in the frontend, or using the REST API.
Based on those parameters, equally-sized bins are created.
For that purpose, the range-to-prefix conversion is applied for every bin in the configured range.
The computational overhead of the conversion algorithm is negligible.
Afterward, all \ac{MAT} entries are written to the data plane in a single gRPC call to reduce configuration overhead.
The configuration of \num{10000} entries takes approximately \qty{100}{\ms}~\cite{ZiFl25} and is therefore negligible.

During traffic generation, the control plane continuously reads the counters of the histogram \acs{MAT}.
Each ternary entry includes a $binIndex$ as action data.
This index is ignored in the data plane but used by the control plane to aggregate counters across entries belonging to the same bin.
The aggregated histogram data is then exported via the REST API.

\subsection{Improved RTT Statistics via Histograms}
In the current version of P4TG, the mean and standard deviation of the \acs{RTT} are computed in the control plane based on sampled values from the data plane. 
This approach can introduce inaccuracies as it relies on a subset of the observed RTT values.
With the introduction of the histogram feature, every packet is counted in exactly one bin, enabling more accurate RTT metric analysis.
Therefore, sampling bias is eliminated.
However, the accuracy of the derived metrics now depends on the histogram configuration, particularly on the number and width of the bins.
The control plane computes the mean and standard deviation of the \acs{RTT} using the midpoint of each bin and the corresponding packet count.
Further, the control plane calculates percentiles from the histogram distribution.
Currently, the $25^{th}$, $50^{th}$, $75^{th}$, and $90^{th}$ percentiles are calculated.
The calculation of additional percentiles will also be possible in the future.

\section{Evaluation}
\label{sec:eval}
In this section, we first evaluate the number of ternary entries required per bin as this may pose a hardware limitation.
Then, we demonstrate the histogram-based RTT measurement using a traffic stream with a log-normal distributed RTT.

\subsection{Approximated Number of Ternary Entries per Bin}
The range-to-prefix conversion algorithm represents each bin using multiple ternary entries.
Let $W$ be the bin width in bit.
According to Gupta et al.~\cite{GuMc01}, a $W$-bit range can be represented by at most $2\cdot W-2$ ternary entries using a range-to-prefix conversion algorithm.
Based on this, the total number for a histogram with $N$ bins (each with a $W$-bit width) is $N \cdot (2\cdot W - 2)$ in the worst case.
In practice, however, due to favorable alignment, the actual number is often lower.
For example, if the width of the range is a power of two and the starting value is aligned to that power, only a single ternary entry is required for a bin.

The field of range-to-prefix conversion has been well researched in the past years.
Many works propose algorithms to reduce the number of required ternary entries in a range-to-prefix conversion~\cite{SuKi10, BrHa18, SuKi10_2, DoSu6}.
They may be explored in the future to optimize the number of ternary entries per bin in P4TG.
However, since the histogram \acs{MAT} in P4TG has a size of \qty{8196}{entries}, the current approach is sufficient to model hundreds of bins.
This size can be increased if required.

\subsection{Demonstrating Histogram-Based RTT Measurement}
In this section, we demonstrate the P4TG RTT histogram feature.
We use P4TG to generate a \ac{CBR} traffic stream with \qty{1518}{byte} frames for approximately \qty{35}{\minute} at a rate of \qty{20}{\gbps}.
The traffic is forwarded through a network that adds a log-normal distributed delay with a mean of \qty{50}{ms} and standard deviation of \qty{1}{ms} to each packet\footnote{Histogram data is collected at a line rate of \qty{400}{\gbps}. However, the network delay emulator is currently most accurate up to \qty{20}{\gbps}.}.
Then, the traffic is sent back to P4TG for \acs{RTT} measurement.
The RTT histogram is configured with a range of $[\qty{46}{\milli\second}, \qty{54}{\milli\second}]$, and \qty{500}{bins}, i.e., a bin width of \qty{20}{\micro\second}.
\fig{histogram} visualizes the measured RTT histogram and the theoretical log-normal distribution.

\figeps[0.85\columnwidth]{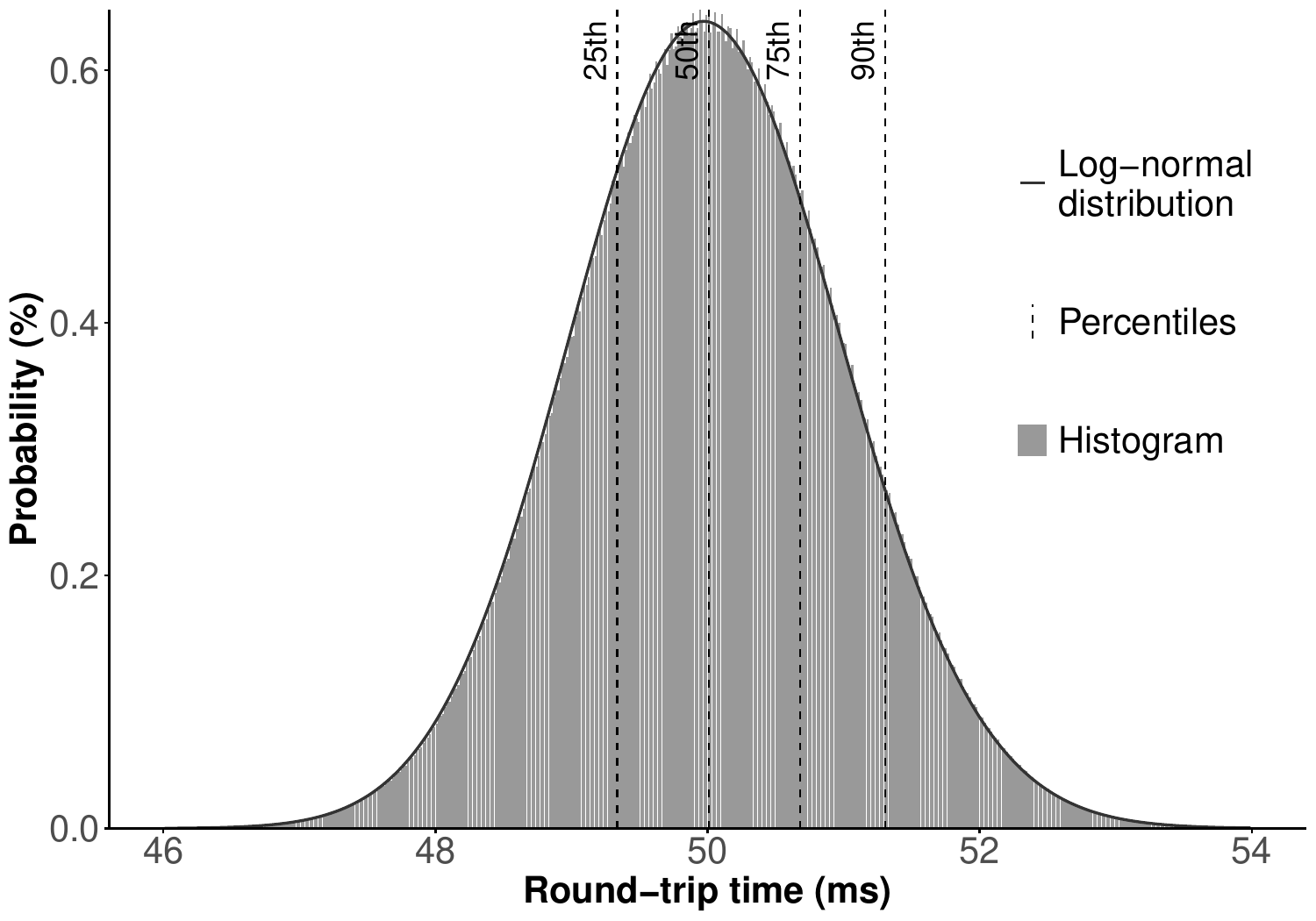}{
RTT histogram with \qty{500}{bins}, ranging from \qty{46}{ms} to \qty{54}{ms}, percentiles and the theoretical log-normal distribution.
}

A total of \qty{3.46}{billion\:packets} was counted.
The control plane calculated $\mu(RTT) = \qty{50.01}{\milli\second}$ and $\sigma(RTT) = \qty{993.31}{\micro\second}$ from the histogram data which matches the configured values.
Further, P4TG calculates the percentiles presented in \fig{histogram}.
It is visible that the histogram matches the theoretical log-normal distribution closely.
This histogram configuration required a total of $\qty{7477}{ternary\:entries}$.

\bibliography{bibliography/literature}

\bibliographystyle{unsrt2authabbrvpp}

\end{document}